\documentclass[aps,prl,twocolumn,showpacs,amsmath,amssymb,groupedaddress,10pt]{revtex4-1}
\usepackage{graphicx,type1cm,eso-pic,color}

\usepackage{graphicx}
\usepackage{dcolumn}
\usepackage{bm}
\usepackage{color}
\newcommand{\sech}{\mathrm{sech} \,}
\begin{document}


\title{Switching chaos on/off in Duffing oscillator}
\author{R. Chabreyrie}
\email{rchabrey@andrew.cmu.edu}
\homepage{http://www.andrew.cmu.edu/user/rchabrey/}
\affiliation{ Mechanical Engineering Department,
Carnegie Mellon University, 
Pittsburgh, Pennsylvania, U.S.A.}

\author{N. Aubry}
\affiliation{ Mechanical Engineering Department,
Carnegie Mellon University, 
Pittsburgh, Pennsylvania, U.S.A.}


\begin{abstract}
Periodic forcing of nonlinear oscillators  generates a rich and complex variety of behaviors, ranging from regular to chaotic behavior. In this work we seek to control, i.e., either suppress or generate, the chaotic behavior of a classical  reference example in books or introductory articles, the Duffing oscillator. 
For this purpose, we propose an elegant strategy consisting of simply adjusting the shape of the time-dependent forcing.
The efficiency of the proposed strategy is shown analytically, numerically and experimentally. In addition due to its simplicity and low cost such a work could easily be turned into an excellent teaching tool. 
\end{abstract}
\maketitle

Almost 100 years ago, as a pioneer, G. Duffing  proposed a periodic forced  nonlinear oscillator \cite{Duffing:18} and with it began the idea that simple systems can display an extraordinary rich and complex dynamic. Today, this key model has a vast number of applications in science and engineering \cite{BookDuffing}.  
Forced nonlinear oscillators displaying chaotic behavior have been observed ubiquitously across a broad range of phenomena, including
fluid devices \cite{Bush:09}, atomic systems \cite{BookAtomic}, 
cardiac tissues \cite{Glass:83}, chemical reactions \cite{BookChaosChem}.
For some applications, the rich and complex chaotic dynamics are desired (e.g., fluid mixing  or data encryption), while for others such dynamics can present major drawbacks and significantly decrease performance (e.g., permanent magnet synchronous motor \cite{Li:98}). 
During the past half century, chaotic phenomena have been under intense investigation. Such booming interest has led to powerful tools that today enable us to elaborate control strategies for such rich and complex phenomena.
More precisely, these strategies use tools such as K.A.M theory, resonance phenomena and linear stability of periodic orbits in order to eliminate \cite{Lima:1990,Guido02,Tounsia:2006}, confine \cite{ChabreyriePRE08} (at a prescribed location and for a specific size), or spread chaos \cite{ChabreyriePhysFluids}. 
In this paper, we present a facile control strategy that enables us to turn on/off the chaotic behavior of the Duffing oscillator.  
Our strategy consists of simply and only adjusting the shape of the time-dependent forcing to either put the system in a chaotic aperiodic or regular periodic oscillatory regime.

\begin{figure}[t]
\includegraphics[scale=.60]{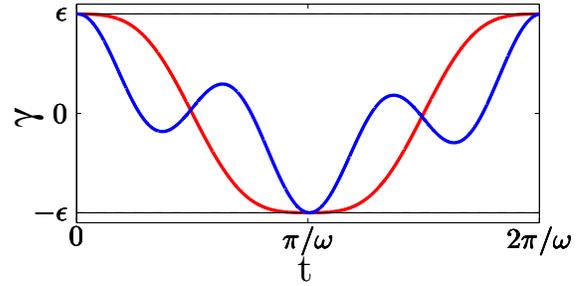}
\caption{\label{fig_1} Two characteristic shapes of the bi-chromic forcing with the pairs of parameters $\kappa_1=\left\{\epsilon_m=.2,\phi=\pi\right\}$ and $\kappa_0=\left\{\epsilon_m=.9,\phi=.35\right\}$.}
\end{figure}

The dynamical system considered here is the classical Duffing's equation \cite{Duffing:18} which is  one of the most common examples in chaotic oscillation teaching texts and research articles. 
Indeed, such a dynamical system externalizes all key characteristics of chaotic system including manifold intersections, strange attractor, positive Lyapunov exponent, etc... (see \cite{Guckenheimer:1983} for a detailed introduction to this system). The equation describes the motion of a damped oscillator with a nonlinear restoring force.  This mechanical system models a spring pendulum  with a nonlinear spring stiffness, the bending angle $\theta$ is then  governed by:
\begin{equation}
\ddot{\theta}=\alpha\theta + \beta\theta^3  -\delta\dot{\theta} + \gamma\left(t;\epsilon,\omega,\epsilon_m,m,\phi\right),
\label{Eq:PhysicalModel}
\end{equation} 
where $\delta$ represents the viscous damping coefficient  $(\alpha,\beta)$, the linear and cubic coefficient in the torsional restoring force.
The ingenuity of this work lies in the time-periodic forcing $\gamma\left(t;\epsilon,\omega,\epsilon_m,m,\phi\right)$
which is a bi-chromic function of the form 
$$
\gamma\left(t;\epsilon,\omega,\epsilon_m,m,\phi\right)=\frac{\epsilon}{N\left(\phi\right)}\left(\cos\left(\omega t\right)+\epsilon_m\cos\left(m\omega t + \phi\right)\right),
$$
where $N$ is a normalizing function such that $-\epsilon\leq\gamma\leq\epsilon$ (see Fig.\ref{fig_1}).\\ 
The strategy consists of simply and only adjusting the shape of the forcing by varying a pair of parameters $\kappa=\left\{\epsilon_m,\phi\right\}$ i.e., the amplitude ratio and frequency while the frequency ratio is maintained fixed at $m=3$  (see Fig.\ref{fig_1}) in order to turn on/off chaotic behavior.
\\

\begin{figure}[t]
\includegraphics[scale=.60]{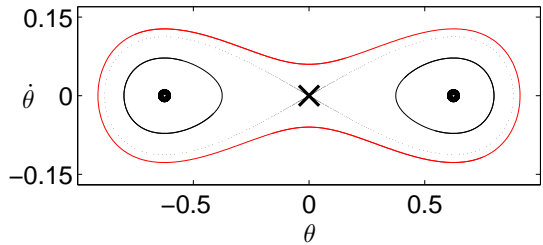}
\caption{\label{fig_0}  Phase space in the unperturbed case i.e, $\delta=0$, $\gamma=0$ with one hyperbolic unstable (cross) and two twin elliptic stable (dot) equilibrium points. The  characteristic separatrix (dotted black line),  partial  (solid blue line) and complete (solid red line) oscillations are represented.}
\end{figure}

\begin{figure}[t]
\includegraphics[scale=.60]{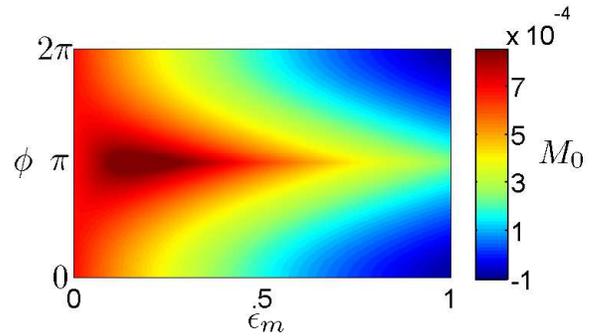}
\caption{\label{fig_2}  $M_0$ versus the pair of  parameters controlling the forcing shape i.e., $\kappa=\left\{\epsilon_m,\phi\right\}$ .}
\end{figure}

\begin{figure}[t]
\includegraphics[scale=.60]{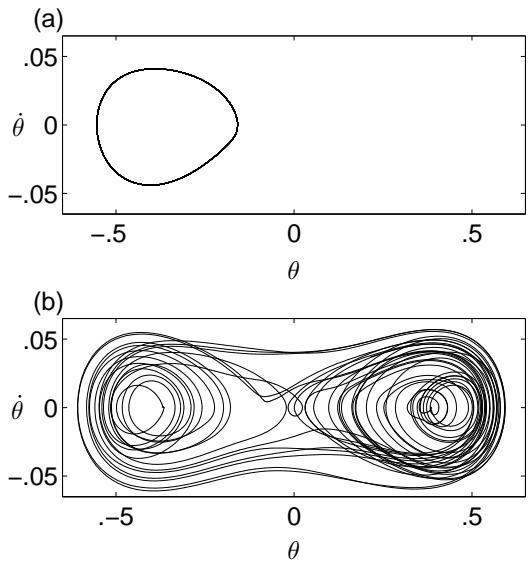}
\caption{\label{fig_3} Phase space from numerical simulation when chaos is turned on (i.e., $\epsilon_m=.2$, $\phi=\pi$) and off (i.e., $\epsilon_m=.9$, $\phi=.35$)  (a)-(b), respectively.}
\end{figure}

\begin{figure}[t]
\includegraphics[scale=.50]{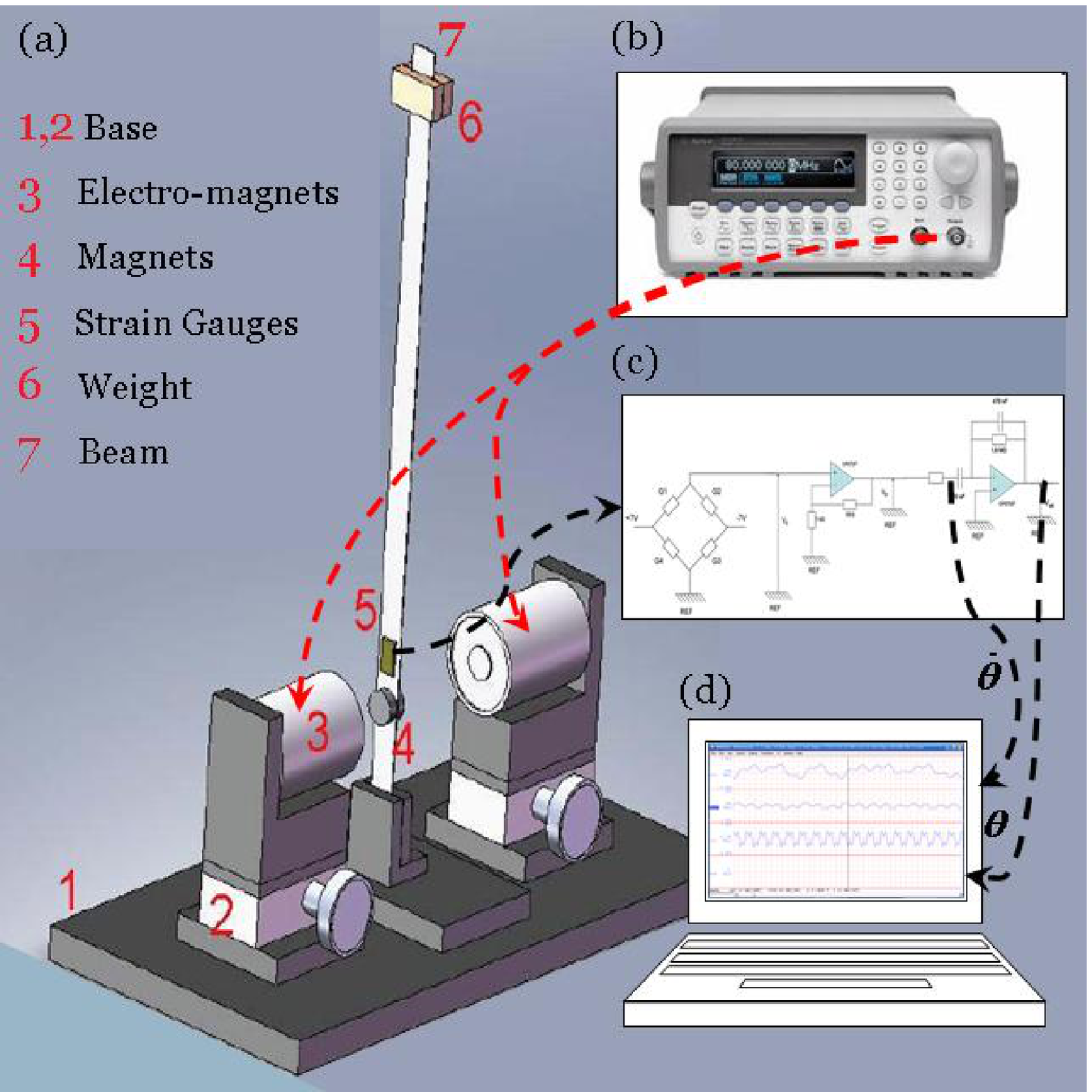}
\caption{\label{fig_4b} 
Sketch of experimental setup.\\
(a) Mechanical oscillator with weight, steel beam \cite{Ruler}, strain gauges \cite{Strain} and electromagnets \cite{ElectroMagnet} from top to bottom.\\
(b) Arbitrary waive generator \cite{AWG}.\\
(c) Sketch of the differential amplifier electrical \cite{Circuit}.\\
(d) Data acquisition \cite{DataQ} and signal treatment system.}
\end{figure}

 As a first step, it is logical to look at the unperturbed unforced case i.e $\delta=0$, $\gamma=0$. Although this case is trivial  it is an excellent groundwork to investigate the system studied. 
In this case the phase space $(\theta,\dot{\theta})$ is characterized by three, one hyperbolic unstable and two twin elliptic stable, equilibrium points Fig.~\ref{fig_0}. Almost all orbits are closed curves showing partial periodic oscillations when initial energy is low (concentric orbits around elliptic points) or complete oscillations when initial energy overcomes a certain threshold (orbits around the pair of elliptic and hyperbolic points). This threshold or separatrix corresponds to the non-closed homoclinic orbits that links the stable and unstable direction of the hyperbolic point. 
In our approach to see how the strategy works, the separatrix and more precisely the stable and unstable manifolds that composed it play a crucial role. Indeed, the ability to turn on/off chaos  by adjusting the forcing shape (i.e., by adjusting the parameters $\epsilon_m$ and $\phi$) can be seen analytically via Melnikov approach. Indeed, a necessary condition for chaos is that the stable and the unstable manifolds of the corresponding hyperbolic fixed point intersect transversally. 
One way to verify this condition is to look at the simple zeros of the Melnikov function which corresponds to  the distance between the stable and unstable manifolds (order of the perturbation) measured along a direction that is perpendicular to the unperturbed homoclinic orbit. In our case, the existence (or nonexistence) of zeros in the Melnikov function can adroitly be seen with the sign of 
\begin{equation} M_0=\max_{t_o}\left|F\left(t_0;\alpha,\epsilon,\omega,\epsilon_m,m,\phi\right)\right|-\left|R\left(\alpha,\beta,\delta\right)\right|,\label{Melnikov_function}
\end{equation} 
\begin{equation}\nonumber
\begin{split}
\mbox{where, }&F\left(t_0;\alpha,\epsilon,\omega,\epsilon_m,m,\phi\right)=\pi\omega\epsilon/N\left(\phi\right)\times\\
&\left[\sech\left(\omega\pi/2\sqrt{\alpha}\right)\sin\left(\omega t_{0}\right)\cdots\right.\\
&\left.\cdots + m\epsilon_{m}\sech\left(m\omega\pi/2\sqrt{\alpha}\right)\sin\left(m\omega t_{0}+\phi\right)\right]
\end{split}
\end{equation} is a periodic function corresponding to the forcing and,
$R\left(\alpha,\beta,\delta\right)=4\alpha^{3/2}\delta/3\sqrt{2\beta}$ represents the unforced oscillator's threshold for the possibility of chaos. In the case where $M_0>0$ the stable and unstable manifolds intersect transversally turning on the possibility of chaos. In the other case i.e., $M_0<0$ the manifolds do not intersect and the possibility of chaos is turned off.\\
Figure \ref{fig_2} illustrates $M_0$ vs. the forcing shape parameters i.e., $\kappa=\left\{\epsilon_m,\phi\right\}$ while all the other parameters are maintained  fixed with values corresponding to the experiment Fig.~\ref{fig_4b} (i.e., $\alpha=.028$, $\beta=-.166$, $\delta=.020$, $\epsilon=.0022$ and $\omega=.20$). With Fig.~\ref{fig_2} we clearly see that by adjusting parameters $\epsilon_m$ and $\phi$ we are able to turn  on the  possibility of chaos (from very cold to hot color) and vice-versa, turn off the possibility of chaos (from hot to very cold color). 
\\
\\
From these analytical results, based on Melnikov approach and illustrated by $M_0$ (see Fig.~\ref{fig_2}), two extreme regions  in the parameter space controlling the  forcing shape can be distinguished: one very hot around $\kappa_1=\left\{\epsilon_m=.2,\phi=\pi\right\}$, and one very cold around $\kappa_0=\left\{\epsilon_m=.9,\phi=.35\right\}$.
Figure \ref{fig_3} qualitatively shows  the ability to turn on/off chaos by numerically integrating the phase space evolution of the system Eq.~\ref{Eq:PhysicalModel} for the pairs of parameters $\kappa_{1/0}$.\\ 
On one hand, when chaos is turned on (i.e., parameters controlling the forcing shape are taken as $\kappa_1$), the system  oscillates by wandering unpredictably in a large domain of the phase space without any apparent pattern (see Fig.~\ref{fig_3}(a)).  On the other hand, when chaos is turned off (i.e., parameters controlling the forcing shape are taken as $\kappa_0$), the system oscillates by staying trapped on a closed curve of the phase space with a periodic pattern (see Fig.~\ref{fig_3}(b)). 
\\

\begin{figure}
\includegraphics[scale=.60]{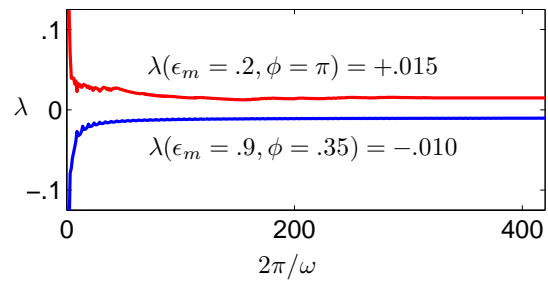}
\caption{\label{fig_4} Leading Lyapunov exponent $\lambda$ vs time $t$ from numerical simulation when is turned on (i.e., $\epsilon_m=.2$, $\phi=\pi$) and off (i.e., $\epsilon_m=.9$, $\phi=.35$)  (a)-(b), respectively.}
\end{figure}

\begin{figure}[t]
\includegraphics[scale=.60]{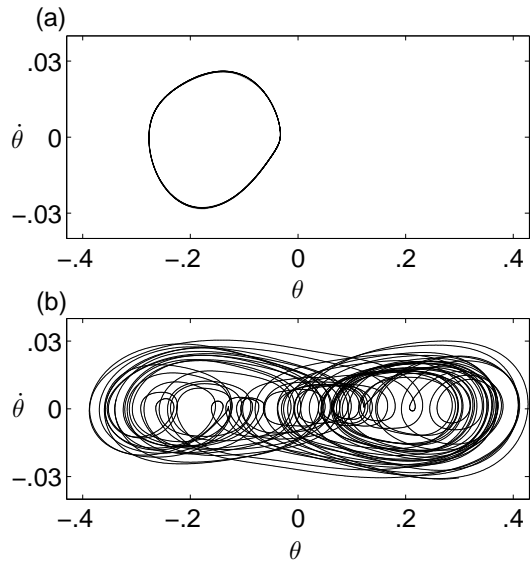}
\caption{\label{fig_5} Nondimensionalized phase space from experiment when chaos is turned on (i.e., $\epsilon_m=.2$, $\phi=\pi$) and off (i.e., $\epsilon_m=.9$, $\phi=.35$)  (a)-(b), respectively.
The nondimensionalized  experimental parameter values  are the same as in Fig. \ref{fig_3}. These nondimensionalized values correspond to: a forcing frequency of $18$ mHz  and an amplitude of $1.25$ Vpp on the magnets; a torsional restoring force cubic and linear  coefficient of $2.5$ and $15.1$ mN.m, and a torsional viscous  damping coefficient of  $.31$ mN.m.s.}
\end{figure}

In order to determine if a system is chaotic and quantify its chaoticity it is important to study the evolution of a cloud of neighboring initial conditions. More precisely, we want to quantify the sensitivity to initial conditions which is a key ingredient in chaotic systems. This sensitivity to initial conditions can be quantified as: 
$$
\left|\delta{\bf X}\left(t\right)\right|\approx e^{\lambda t}\left|\delta{\bf X}\left(0\right)\right|,
$$
where $\delta{\bf X}\left(t\right)$ is the evolution of a close cloud of neighboring initial conditions and $\lambda$ is the mean rate of separation of the system called Lyapunov exponent. When $\lambda>0$, any two orbits that start out very close to each other diverge quickly, otherwise when $\lambda<0$ all points in the neighborhood converge toward the same periodic orbit. In order to compute the Lyapunov exponent we first consider the time evolution of the Jacobian  $J^t\left(\theta,\dot{\theta}\right)$ given by the tangent flow and the matrix of variations as
$$
\frac{dJ^t}{dt}=
\left(\begin{array}{cc}
1&0\\
\frac{\partial \ddot{\theta}}{\partial\theta}&\frac{\partial\ddot{\theta}}{\partial\dot{\theta}}
\end{array}
\right)J^t,
$$
where  $J^0=I$ is the two-dimensional identity matrix. 
Then, the leading Lyapunov exponent for a finite-time $t=\tau$ is defined as $\lambda\left(\tau\right)=\frac{1}{\tau}\ln\left|\Lambda_{max}\right|$
where $\Lambda_{max}$ is the largest  eigenvalue (in norm) of the Jacobian $J^{\tau}$.\\ 
Figure \ref{fig_4} shows the leading Lyapunov exponent $\lambda$ vs. time $t$ when chaos is turned on/off (i.e., with forcing shape parameters $\kappa_{1\slash0}$) . On one hand, when chaos is turned on Fig.~\ref{fig_4} (a) displays a Lyapunov exponent converging toward a positive value of $\lambda=.015$. Such a value indicates high sensitivity to initial conditions in the system and consequently high chaotic behavior. In the other hand, when chaos is turned off, Fig.~\ref{fig_4} (b) shows a negative Lyapunov exponent value  indicating no sensitivity to initial conditions and consequently no chaotic behavior. All orbits converge toward the periodic orbits shown in Fig.~\ref{fig_3}(a).
After having been applied analytically and numerically, this strategy for turning on/off chaos has also been implemented experimentally.
The design and construction of our experimental system based on \cite{Berger:97} and is illustrated in Fig.~\ref{fig_4b}. Such a setup presents the advantage of being economical and  simple to realize while still being able to be qualitatively described by Eq. (\ref{Eq:PhysicalModel}).  
The oscillator consists of a vertical steel beam with a brass weight clamped at its top.  This beam is subjected to a forcing produced by a pair of electromagnets connected to an arbitrary wave generator.
This generator fed the electromagnets with a bi-chromic voltage wave identical to $\gamma$ (see Fig.~\ref{fig_1}). In addition, four strain gauges  glued on the beam send an electrical signal to an differential amplifier electrical circuit providing voltages proportional to the bending angle from the vertical $\theta$ and the  angular velocity $\dot{\theta}$. 
The experimental procedure consists of starting  the oscillation around $(\theta_0=0,\dot{\theta}_0=0)$, then waiting $20$ min (i.e., more than $200$ times the forcing period) in order to pass the transient mode to make sure the system has well converged toward its attractor, and finally recording the data at $60~\mbox{Sa}.\mbox{s}^{-1}$ for $20$ min.  
Figure~\ref{fig_5} shows experimental results for the phase space evolution of our experimental Duffing oscillator Fig.~\ref{fig_4b}. 
These results verify the strategy proposed i.e., by simply and only adjusting the forcing shape we are able to turn on/off chaos. As predicted by the analytical and numerical results, chaos is turned on/off when the forcing shape parameters are adjusted as $\kappa_{1\slash 0}$ (see Fig. \ref{fig_1} for illustration). 
On one hand, when chaos is turned on, the system oscillates in unpredictable fashion in a large part of the phase space domain. On the other hand, when chaos is turned off, the system oscillates in a completely predictable periodic pattern.
We have shown analytically, numerically and experimentally that chaotic behavior of the famous Duffing oscillator can be controlled successfully by  the bi-chromic forcing $\gamma$. More precisely, by adjusting the forcing shape (i.e., the amplitude ratio $\epsilon_m$ and the phase difference $\phi$ between the two harmonics composing the forcing) chaotic behavior can be turned on/off. 
This control strategy has the great advantage of being extremely easy to set up since neither feed back loop nor addition of time variation input are necessary. Furthermore, this strategy does not require important investments such as increase in force amplitude or frequency. \\ 
Due to the universality of the Duffing equation in science, we expect this strategy  to be applicable to other oscillatory experimental systems as well.
\begin{acknowledgments}
The authors are grateful to C. Chandre for his help and suggestions, and thanks  M. Labour, N. Zaharia, B. Belda and H. Wang for their assistance in performing the experiments.
\end{acknowledgments}

\bibliography{biblioII}

\end{document}